\documentclass[twocolumn,showpacs,amsmath,amssymb,aps,prl]{revtex4}
\usepackage{graphicx,epsfig}% Include figure files
\usepackage{dcolumn}% Align table columns on decimal point
\usepackage{bm}
\begin{document}
\title{
Large-$n$ Excitations in the Ferromagnetic Ising Field Theory in a Small  Magnetic Field: 
Mass Spectrum and Decay Widths}
%\date{\today}
\author{S. B. Rutkevich}
\email{rut@ifttp.bas-net.by}
\affiliation{ Institute of Solid State and Semiconductor Physics,
220072 Minsk, Belarus.}
\begin{abstract}
In presence of a small magnetic field $h$, the elementary excitations in the scaling 
two-dimensional Ising model  are studied perturbatively in $h$ in the ferromagnetic phase.
For excitations with large numbers $n$, the mass spectrum is obtained in the first order in $h$.
The decay widths of  excitations with energies above the stability threshold are calculated in the leading 
$h^3$-order. 
\end{abstract}
\maketitle

 During the past decades there has been much progress
in  study of the Ising Field Theory (IFT). For a recent review, see \cite{Del04}.  Being the 
field-theoretical realization of the  two-dimensional 
Ising model in the scaling limit, IFT provides  a direct information about the 
2D Ising universality class. IFT can describe also dynamical properties of many one-dimensional 
condense-matter systems, the best known example is the one-dimensional Ising magnet  at 
$T=0$ near the quantum 
phase transition point \cite{Sach99,Tsv04}. Viewed as a particle theory, IFT 
turns into a 
beautiful model of quark confinement in which new ideas and methods
 can be efficiently worked out \cite{McCoy78, McCoy95,FonZam2003,Tsv04}.  

Physics of IFT is determined by the 
scaling combination $\eta=m/| h|^{8/15}$ of parameters $m$ and $h$, which are proportional 
to the deviations of 
temperature  and magnetic field  in the  2D Ising model from their critical point values 
$T=T_c$ and $H=0$. 
Being not integrable for generic $\eta$, the IFT has exact solutions 
only  at $\eta=\pm\infty$ and $\eta=0$. 
The former case is free-fermionic and corresponds to the Onsager's 
exact solution 
\cite{Ons44} of the lattice Ising model in zero field. Solution of 
IFT at  $\eta=0$, that  describes the scaling Ising model 
at  the critical temperature  
in a magnetic field, was found by Zamolodchikov \cite{ZamH}. Beyond the integrable 
cases, the main progress has been achieved  by exploiting  perturbation expansions 
around  $\eta=0$  and $\eta=\pm\infty$  \cite{McCoy78,Del96,Fat98,FonZam2003,FZWard03,Del05}. 
In the present paper we focus on the perturbative study of the Ising field theory
in the limit $\eta\to+\infty$, i.e. at $m>0$ and $h\to0$. This case corresponds to the
 ferromagnetic (low temperature) small-field scaling regime of the 2D Ising model. 

At $h=0$, $m>0$, the particle sector  of the IFT
contains one free fermion  of the mass $m$,  which can be interpreted as a 
domain wall (kink) interpolating
 between the two degenerate vacua. 
External  magnetic field $h>0$ removes degeneration, and induces 
a long-range confining interaction
between fermions (kinks). As the result, an isolated fermion gains an 
infinite energy at a whenever small field $h$, and the kink-antikink bound states
become the single-particle excitations of the model. This makes a link with  the 
quark confinement physics. 
 Accordingly, in what follows the fermions and their bound states are referred to as "quarks"
and "mesons", respectively.
At a small external field, the weak confinement regime is realized: the meson 
masses $M_n$ fill dense the interval 
above $2 m$, the mesons with $M_n>2 M_1$ are unstable, where $M_1$ is the lightest meson mass, 
$M_1\approx 2 m$. 

Though the  qualitative picture outlined above was understood already 
in 1978 \cite{McCoy78}, only few quantitative results on the weak confinement in  
IFT have been obtained ever since. 
The initial part 
of the meson spectrum $M_n$ in the region 
\begin{equation}
M_n-2 m\ll m  \label{smM}
\end{equation}
was first obtained by McCoy and Wu \cite{McCoy78}
in the non-relativistic approximation. Relativistic corrections to this spectrum in the 
same region (\ref{smM}) 
have been recently found 
by Fonseca and Zamolodchikov \cite{FonZam2003,FZWard03}. 
In the present paper we calculate perturbativly in $h$ the meson masses $M_n$ 
for large $n\gg1$, 
as  well as the decay widths  of the nonstable mesons.
 
Ising field theory is determined  by the  euclidean action describing  massive 
interacting Majorana 
fermions in the plane $ \mathbb{R}^2$. For  its  explicit expression see for example 
Refs. \cite{Del04,FonZam2003}. 
The corresponding hamiltonian  can be written in the form
\begin{eqnarray}
\mathcal{H}&=& \int_{-\infty}^\infty \frac{dp}{2 \pi} \,\omega(p)\, a^{\dagger}_p\, a_p  +
V , \label{Ham} \\
\textrm{where  }\;\;\;V&=&-h\int_{-\infty}^\infty  d\rm{x}\,\sigma(\rm{x}),   \label{V}
\end{eqnarray}
and  
$\omega(p)=\sqrt{p^2+m^2}$ is the spectrum of free fermions. 
Fermionic operators $a^\dagger_p, \,a_p$  creating and annihilating a bare quark with 
a momentum 
$p$ obey the canonical 
anticommutational relations
\begin{equation}
\{ a_p ,a^\dagger_{p' }\} =2 \pi \,\delta(p-p'), \; 
 \{ a_p ,a_{p'} \} = \{ a^\dagger_p ,a^\dagger_{p'} \}= 0.  \nonumber
\end{equation}
Notations $|p_1...p_N \rangle$ = $a^\dagger_{p_1} ...a^\dagger_{p_N}|0\rangle$ 
and $\langle p_1...p_N|=\langle 0|  a_{p_1} ...a_{p_N}$
will be used.
The order spin operator $\sigma({\rm x})$ can be determined in the infinite line 
${\rm x}\in \mathbb{R}$ as a normally ordered 
exponent \cite{Jimb80,Rut01}:
\begin{subequations}
\label{alleq}
{\setlength\arraycolsep{0.5pt}
\begin{eqnarray}
 \sigma ({\rm x})&=&\bar{\sigma} : e^{\rho({\rm x)}/2} :\;, \label{sigma}\\
\frac{ \rho({\rm x})}{2}&=&  \int_{\rm x}^\infty d{\rm x}' \big(\chi({\rm x}',{\rm y})\,
\partial_{\rm y} \chi ({\rm x}',{\rm y})\big){\big |}_{{\rm y}=0},            \label{rox}   \\
\chi({\rm x},{\rm y})=&i& \int_{-\infty}^\infty \frac{dp}{2 \pi} 
\frac{e^{i p x}}{\sqrt{\omega(p)}}\big(a^{\dagger}_{-p} 
e^{ \omega(p){\rm y}}-a_p e^{- \omega(p){\rm y}}\big),  \nonumber
\end{eqnarray}}
\end{subequations}
where $\bar{\sigma}=m^{1/8}2^{1/12}e^{-1/8}A^{3/2}$ is the zero-field vacuum 
expectation value of the order field (spontaneous magnetization), 
$A=1.28243...$ is Glaisher's constant. It is useful to 
rewrite operator $:\rho( 0)/2:$  in the form
{\setlength\arraycolsep{0.5pt}
\begin{eqnarray} 
&:&\frac{\rho(0)}{2}:=i \iint_{-\infty}^\infty\frac{dp\, dk}
{8\pi^2 \sqrt{\omega(p)\,\omega(k)}} \cdot \label{rhoo}\\   
&\bigg(&(a_p a_k +a^{\dagger}_{p} a^\dagger_{k})\, 
\frac{\omega(k)-\omega(p)}{  p+k}+2 a^\dagger_{p}\,a_k\,
\frac{\omega(k)+\omega(p)}{p-k-i 0}\bigg). \nonumber
 \end{eqnarray}}
One can easily verify, that  Eqs.~(\ref{alleq}), (\ref{rhoo}) lead to the  
well-known formfactors \cite{FonZam2003,Berg79} 
$\langle p_1...p_{N}|\sigma(0)|p'_1...p'_{N'} \rangle $ of the order field in 
the infinite system.  
 Note, that the field theory determined by hamiltonian (\ref{Ham})-(\ref{alleq}) is 
Lorentz invariant.

We are concerned with the problem of the meson mass spectrum 
 in the low-temperature phase 
IFT in the small field regime $h \bar{\sigma}\ll m^2$. 
 The meson
 masses $M_n$ are determined by the
eigenvalues  of the full Hamiltonian (\ref{Ham}):
\begin{equation}
({\mathcal H} -E_{\rm vac})\mid \Phi_n\rangle = M_n \mid \Phi_n\rangle,
\end{equation}
where $E_{\rm vac}$ is the ground-state energy, and $ \mid \Phi_n\rangle$ is a 
one-meson state with zero momentum.
In the two-quark approximation one considers the projection of this equation onto 
the two-fermion  subspace  $\mathcal F_2 $ of the 
Fock space $\mathcal F$:
\begin{equation}
\mathcal P_2 \,\mathcal H  \, \mathcal P_2 \mid 
\Phi_n\rangle = M_n \mid \Phi_n\rangle,  \label{eig2}
\end{equation}
where $\mathcal P_2$ is the orthogonal projector on $\mathcal F_2 $. 
The scalar product of (\ref{eig2}) with the 
bra-vector $\langle -p,p\mid$ yields the  Bethe-Salpeter equation
\begin{eqnarray}
(2\,\omega (p)-M_n)\, \Phi_n (p) = \bar{\sigma}\, h 
\int_{-\infty}^\infty \frac{1}{ \omega(p)\omega(p') } \cdot \label{BS}\\
\Bigg[\bigg(\frac{ \omega(p)+\omega(p') }{p-p'}\bigg)^2  +
\frac{1}{2}\frac{p\,p'}{\omega(p)\omega(p') }\Bigg]
\Phi_n(p')\frac{d p'}{2 \pi},\nonumber
\end{eqnarray}
first obtained by Fonseca and Zamolodchikov \cite{FonZam2003}. Here the principal 
value integral is implied. 
The two-particle wave function 
in the rest frame in the momentum representation is defined via 
\begin{equation}
\Phi_n(p) \equiv R^{-1/2} \langle -p,p\mid \Phi_n\rangle,  \label{wf}
\end{equation}
where $R$ is the system length in the x-direction, which is put to infinity in the 
final formulas.
The function  $\Phi_n (p)$ is odd $\Phi_n(-p)=-\Phi_n(p)$, and should be 
normalized as
\begin{subequations}
\begin{eqnarray}
 \int_0^\infty \frac{dp}{2 \pi} \mid \Phi_n (p)\mid^2&=& 1,   \label{norm} \\
 \textrm{if        }  \langle \Phi_n \mid \Phi_n \rangle &=&1.     \label{nno}
\end{eqnarray}
\end{subequations}

It is worthwhile to rewrite equation (\ref{BS}) in the coordinate representation:
\begin{equation}
(2 \hat{K}- M_n) \phi_n({\rm x}) = -2 \bar{\sigma} 
h |{\rm x}|\,\phi_n({\rm x})+2 \bar{\sigma} h \hat{U} \,\phi_n({\rm x}). \label{BSx}
\end{equation}
Here   $\phi_n({\rm x})$ denotes the configuration-space wave function
\begin{subequations}
\begin{eqnarray} 
\phi_n({\rm x})= \int_{-\infty} ^\infty\frac{d p}{2 \pi} \,e^{i p {\rm x}} 
\Phi_n(p), \label{Fou}\\
\phi_n(-{\rm x})=-\phi_n({\rm x}), \label{odd}
\end{eqnarray}
\end{subequations}
and  the integral operators $\hat{K}$ and  $\hat{U}$ have the kernels
\begin{subequations}
\begin{eqnarray}
K({\rm x},{\rm x}')= \int_{-\infty} ^\infty \frac{d p}{2 \pi} \,
e^{i p ({\rm x}-{\rm x}')} \sqrt{p^2+m^2}, \\
U({\rm x},{\rm x}')=\frac{1}{2} \iint_{-\infty}^\infty \frac{d p\,dp'}{4 \pi^2}\, 
\frac{\exp[i ( p {\rm x} -p' {\rm x}' )]}{ \omega(p)\omega(p') } \cdot  \\
\Bigg[ \bigg( \frac{ \omega(p)-\omega(p') }{p-p'}\bigg)^2  +
\frac{1}{2}\frac{p\,p'}{\omega(p)\omega(p') }\Bigg]. \nonumber
\end{eqnarray}
\end{subequations}
Discrete-lattice versions of equation (\ref{BSx}), without the last term in 
the right-hand side,  were studied in papers \cite{Rut99,Rut01}.

All terms in  (\ref{BSx}) have clear physical meaning. The left-hand side 
contains the relativistic kinetic energy
operator $\hat{K}$. The first term in the right-hand side describes the 
long-range confining attractive potential, 
which linearly increases with the distance between quarks $|{\rm x}|$. 
Operator $2 \bar{\sigma} h  \hat{U} $ corresponds to the short-range 
diagonal interaction
between quarks, which exponentially vanishes at large distances  
$|{\rm x}|\gg m $. 

 At small $h$, the lower part of the spectrum $M_n$ of 
equation (\ref{BSx}) has been  obtained by Fonseca and Zamolodchikov  \cite{FonZam2003}
 in the "non-relativistic" approximation 
\begin{equation}
\frac{M_n-2 m}{m}= \zeta^{2/3} z_n - \frac{\zeta^{4/3}}{20} z_n^2 +  
\zeta^{2} \Big(\frac{11\,z_n^3}{1400}-\frac{57}{280}\Big)+O(\zeta^{8/3})  \label{mZ}
\end{equation}
where $\zeta$ is a dimentionless parameter proportional to $h$: $\zeta=2\bar{\sigma} h/m^2$,  
and $-z_n, n=1,2,...$
are zeros of the Airy function, ${\rm Ai}(-z_n)=0$. The leading term of  expansion (\ref{mZ}) 
reproduces the old result of  McCoy and Wu \cite{McCoy78}.

It is worth stressing, that expansion ({\ref{mZ}) describes the spectrum 
of equation (\ref{BSx}), only if  
{\it two conditions} are satisfied together:
 $\zeta\ll 1$ and $n \ll \zeta^{-1} $.  
Exploiting (\ref{mZ}) and the large-$n$ 
expansion for zeros of the Airy function \cite{AbrSt} 
\begin{equation}
z_n=b_n \Big[1+ \frac{5}{48} b_n^{-3} -\frac{5}{36} b_n^{-6} + O(b_n^{-9}) \Big] 
\end{equation}
with $b_n=\Big[\frac{3 \pi}{8}(4n-1)\Big]^{2/3}$, one can see, that  
inequality $n \ll \zeta^{-1}$ is equivalent to (\ref{smM}).
Condition  (\ref{smM}) guarantees, that the wave function $\phi_n({\rm x})$
slowly varies on the scale $\Delta {\rm x}\sim m^{-1}$.  This
allowed Fonseca and Zamolodchikov  to replace the relativistic kinetic energy operator
$ \hat{K} $  by its non-relativistic counterpart,
and to reduce integral equation (\ref{BSx}) to the (differential) perturbed Airy  equation.  

We obtain the mass spectrum of equation (\ref{BSx}) in a different, large-$n$ region, 
namely at
$ \zeta\ll 1,    \; n \gg 1$.
The resulting expansion  reads as
\begin{equation}
M_n = m\,u_n +M_n^{(2)}  +O(\zeta^3),  \label{eM}
\end{equation}
where $ u_n $  solves   equation
\begin{equation}
\frac{u_n}{2} \bigg( \frac{u_n^2}{4} -1 \bigg)^{1/2} -{\rm arccosh}(u_n/2)= 
\bigg( n-\frac{1}{4}  \bigg) \,\pi \,\zeta, \label{un}
\end{equation}
and
\begin{equation}
  M_n^{(2)} = \frac{\zeta^2 \,m}{u_n^3}\bigg[ \frac{40}{3(u_n^2-4)^2}+
 \frac{6}{u_n^2-4}-1  \bigg].  \label{2M}
\end{equation}
Two notes are in order.
\newline
(i) While the small-$n$
expansion (\ref{mZ}) is not analytical in $\zeta$  at fixed $n$, the large-$n$ 
expansion (\ref{eM})-(\ref{2M})
is analytical in $\zeta$, but  at fixed $n\,\zeta={\rm const}$. 
\newline 
(ii) The small-$n$ and  large-$n$ regions 
cross at $1\ll n \ll 
\zeta^{-1}$, covering together all possible 
values of $n\in\mathbb{N}$.
In the crossover region $1\ll n \ll \zeta^{-1}$, the both small-$n$ and 
 the large-$n$ expansions (\ref{mZ}) and (\ref{eM})-(\ref{2M})
 can be reduced to the same form:
\begin{widetext}
\begin{equation}
\frac{M_n-2 m}{m}= \zeta^{2/3} \bigg[b_n+\frac{5}{48 }b_n^{-2}+O(b_n^{-5})\bigg]+
\zeta^{4/3} \bigg[ -\frac{b_n^{2}}{20 } -\frac{b_n^{-1}}{96 } +O(b_n^{-4})\bigg]  +
\zeta^{6/3} \bigg[ \frac{11 \,b_n^{3}}{1400 } -\frac{901}{4480} +O(b_n^{-3})\bigg]+
O(\zeta^{8/3}).  \label{cross}
\end{equation}
\end{widetext}

The leading term in expansions (\ref{eM}) describes  semiclassical energy levels, given by the 
Bohr-Sommerfeld quantization rule, as one should expect for $n\gg1$. In our  problem,
the latter can be  reformulated  in a relativistic covariant form.  

Consider   two interacting particles with coordinates 
${\rm x}_1(t), \, {\rm x}_2 (t)\in \mathbb{R} $ 
described by the classical action 
\begin{subequations}
\begin{equation}
{\mathcal A}_{\rm cl} =- m \int_0^{t_m} dt [ (1-\dot{\rm x}_1^2)^{1/2} + 
(1-\dot{\rm x}_2^2)^{1/2} ] -  2 h \bar{\sigma} S, 
\end{equation}
\vspace{-.5cm}
\begin{equation}  
 \textrm{with   } 
S= \int_0^{t_m} dt\, [ {\rm x}_2 (t) -{\rm x}_1 (t)], 
\end{equation}
\end{subequations}
under the following constrains : $ {\rm x}_1 (0)={\rm x}_2 (0)$,  
${\rm x}_1 (t_m)={\rm x}_2 (t_m)$,  and
$ { \rm x}_1 (t)<{\rm x}_2 (t) $ for $0<t<t_m$.  It is evident, that 
action ${\mathcal A}_{\rm cl}$ is relativistic invariant,
since  $S$ is just the area of the loop
in the plane $({\rm x}, t)$, bounded by the  particle world-paths 
$ { \rm x}_1 (t)$, ${\rm x}_2 (t) $.
Then, the quantization condition 
\begin{equation}
2 h \bar{\sigma} S = 2\pi \bigg( n-\frac{1}{4}  \bigg), \;\;  n\in\mathbb{N}  \label{smq}
\end{equation}
leads to the semiclassical 
mass spectrum  $M_n=m \,u_n$ of the pair, where $u_n$ is determined  by (\ref{un}). Emphasize,
 that the quantization condition  (\ref{smq}) is relativistic covariant, and
 does not imply any constrains on the total momentum of two particles.

Let us briefly explain now, how the large-$n$ expansion (\ref{eM})-(\ref{2M}) was obtained.  
One could not treat
the right-hand side of equation (\ref{BSx}) as a perturbation at a whenever small $h$, since 
the product $h{\rm x} $ in the first term could be large at large enough $ {\rm x} $. 
To overcome this difficulty,  the  strategy familiar from the theory of the 
magnetic breakdown 
in normal metals \cite{LL9} was applied. We divided the ${\rm x}$-axis into 
three intervals by the points
$\pm a$,  where $m^{-1}\ll a\ll (m\,\zeta)^{-1}$, 
 and considered (\ref{BSx}) 
in these regions separately. It is natural to call the  middle interval $-a<{\rm x} <a$  by 
the  "scattering region", and surrounding ones  - by "transport regions". 
In the transport regions, the relative
motion of quarks is semiclassical for  $n\gg1$. 
 In the right transport region $x>a$, the wave function $\phi_n({\rm x} )$  
approaches with exponential accuracy
to the function  $ C \phi_r (x, M_n)$, 
{\setlength\arraycolsep{.0pt}
\begin{eqnarray}
 \phi_r ({\rm x}, M_n)&=& \int_{-\infty}^\infty \frac{dp}{2\pi} 
\exp \bigg[\frac{i\big(f(p)-M_n\, p \big)}{2 h \bar{\sigma}}
 +i p{\rm x}  \bigg], \label{rsol} \\
 \textrm{with  } f(p) &=& \,2 \int_0^p dp' \,\omega (p') ,\nonumber
\end{eqnarray}}
and some constant $C$.
One can easily verify, that function  (\ref{rsol}) solves equation 
\begin{equation}
(2 \hat{K}- M_n) \phi_r({\rm x}, M_n) = -2 \,\bar{\sigma}\, h\, 
{\rm x}\, \phi_r({\rm x},M_n)   \label{righteq}
\end{equation}
in $ \mathbb{R}$. 
In the left  transport region ${\rm x} <-a$, function $\phi_n({\rm x} )$ should 
approach to $-\phi_r (-{\rm x} )$ 
due to the antisymmetric condition (\ref{odd}). 
In the scattering region  
$-a<{\rm x} <a$,  equation (\ref{BSx}) was  solved perturbativly in $h$ to the first order. 
Joining at ${\rm x} \simeq \pm a$    solutions of different regions 
provides us the condition, which 
determines  the discrete mass spectrum (\ref{eM})-(\ref{2M}).

Note, that in the zero order in $h$, the normalized 
(according to (\ref{norm}), (\ref{Fou}))
 wave function 
can be written as 
{\setlength\arraycolsep{.0pt}
\begin{equation}
\phi_n({\rm x})=\sqrt{\frac{\pi}{p_0}}\,
[ \theta({\rm x}) \phi_r ({\rm x},m u_n)- 
\theta(-{\rm x}) \phi_r(-{\rm x},m u_n)],  \label{semic}
\end{equation}}
where $\theta({\rm x})$ is the step-function, and $ p_0= [(M_n/2)^2-m^2]^{1/2}$.  
In the scattering region $|{\rm x}|\lesssim m^{-1}$,  
the wave-function (\ref{semic})
has the asymptotics:
\begin{equation}
\phi_n({\rm x})\simeq  (-1)^{n-1}\, \frac{(m\, u_n h \bar{\sigma})^{1/2}}{p_0}
\sin(p_0\, {\rm x}). \label{smx}
\end{equation}

The above analysis has  been limited by the two-quark approximation (\ref{eig2}). 
However, the exact meson state contains, of course,  multi-quark components,
which should also contribute to the meson mass \cite{FonZam2003}.  Let us 
discuss qualitatively  such multi-quark effects. Again, it is reasonable to consider 
the  
transport and the scattering regions separately. 

In the transport regions $|{\rm x} |\gg m^{-1}$ two quarks move 
independently one from another 
(apart from the long-range attractive force). 
Multi-quark effects here can be absorbed, at least partly, 
by radiative corrections to the parameters 
$m$ and $\bar{\sigma}$ in equation (\ref{BSx}).
Correction $\delta m$ to the quark mass known due to Fonseca and 
Zamolodchikov \cite{FZWard03} 
is of order  
$\delta m\sim h^2$. Simple perturbative estimates show, that the leading correction to 
$\bar{\sigma}$ is of the same $h^2$-order.

In the scattering  region $|{ \rm x} | \lesssim m^{-1} $, two quarks with 
momenta $\pm p_{0}$  collide. 
 This  collision is elastic, if $M_n< 4 m $. In this case virtual multi-quark 
processes lead  to the second-order corrections 
to the scattering phase, which in turn give rise to the meson mass correction in the 
$h^3$-order. Above the two-meson
threshold, nonelastic scattering channels open, causing decay of mesons  with $M_n> 4 m$. 
The meson decay width $\Gamma_n$ 
can be obtained in the leading $h^3$-order from  the  Fermi's golden rule:
\begin{equation}
\Gamma_n= 2 \pi  \sum_{\rm out} 
\mid  \langle  {\rm out}\mid V \mid \Phi_n \rangle \mid^2   \delta (m \,u_n-E_{\rm out}).   
         \label{GFermi}
\end{equation}
Here the  interaction operator $V$ is given by (\ref{V}), an out-state can be taken as 
$\mid  {\rm out}\rangle =    \mid p_1\, ...p_{2j} \rangle$,  and $ \mid \Phi_n \rangle $ 
is the meson state, normalized according to (\ref{nno}),  in the two-quark semiclassical 
approximation.
In equation (\ref{GFermi}) one can substitute
\begin{equation}
 \mid \Phi_n \rangle \to \bigg(\frac{m\, u_n h \bar{\sigma}}{4p_0^2 R}\bigg)^{1/2}
\mid p_0,-p_0 \rangle, \label{rep}
\end{equation}
since quarks collide in the scattering region, 
where asymptotics (\ref{smx}) is applicable.
Then we immediately obtain from (\ref{GFermi}) an explicit representation 
for $\Gamma_n$ in terms 
of the well-known formfactors \cite{FonZam2003,Berg79} 
 of the order operator $\sigma(0)$:
\begin{widetext}
\begin{equation}
\Gamma_n = \frac{{h^3}\bar{\sigma} \,u_n} {m(u_n^2-4)}\sum_{j=2}^{j_m  }
 \frac{1}{(2 j)!} \int_{-\infty}^\infty \frac{d p_1 ... d p_{2 j}}{(2 \pi)^{2j-2}} 
\,\delta (p_1+\cdots+ p_{2 j})\,\delta (\omega_1+\cdots+\omega_{2 j}-m \,u_n) 
\mid \langle p_{2j} ... p_1  \mid \sigma(0)\mid  p_0, -p_0 \rangle\mid^2,  \label{Gam}
\end{equation}
\end{widetext}
where  $\omega_i \equiv \omega (p_i)$,  and $j_m=[u_n/2]$ stays for the 
integer part of $u_n/2$. 
The extra $h$-factor appears in (\ref{Gam})  from (\ref{rep}).  It reflects, 
that quarks spend almost all the time in the transport regions, and only rarely come up
in the scattering region, from which the decay  then occurs.

In summary, we have obtained the large-$n$ excitation spectrum  
in the Ising field theory in the weak confinement regime perturbativly in 
applied field $h$. We argue,  that the two particle approximation
is sufficient to determine the {\it entire spectrum} up to the first order in
$h$, whereas, starting from the second order in $h$, one should take into account 
nondiagonal multi-fermion contributions. For energies above the stability 
threshold, the excitation 
decay width  has been determined in the leading order in $h$.
\begin{acknowledgments}
This work is supported by the Fund of Fundamental Investigations of Republic of Belarus.
\end{acknowledgments}
%\bibliography{BankIsing,mypapers}

\end{document}